\documentstyle[times,pramana,epsf,epsfig,floats,amsmath,axodraw]{ias}

\input paperdef

\begin{document}



\thispagestyle{empty}
\setcounter{page}{0}
\def\thefootnote{\fnsymbol{footnote}}

\begin{flushright}
hep-ph/0611372
\end{flushright}

\vspace{1cm}

\begin{center}

{\large\sc {\bf Electroweak Precision Data and Gravitino Dark Matter}}
\footnote{talk given at the {\em LCWS06}, 9-13 March
  2006, Bangalore, India}

\vspace{1cm}

{\sc S.~Heinemeyer$^{1\,}$%
\footnote{
email: Sven.Heinemeyer@cern.ch
}%
}

\vspace*{1cm}

$^1$ Instituto de Fisica de Cantabria (CSIC--UC), Santander,  Spain 

\end{center}

\vspace*{0.2cm}

\BC {\bf Abstract} \EC
Electroweak precision measurements can provide indirect information about
the possible scale of supersymmetry already at the present level of
accuracy.  
We review present day sensitivities of precision data in mSUGRA-type
models with the gravitino as the lightest supersymmetric particle
(LSP). 
The $\chi^2$~fit is based on $\MW$, $\sweff$, $(g-2)_\mu$, $\br(b \to s \ga)$ 
and the lightest MSSM Higgs boson mass, $\Mh$. 
We find indications for relatively light soft supersymmetry-breaking masses,
offering good prospects for the LHC and the ILC, and in some cases
also for the Tevatron.

\def\thefootnote{\arabic{footnote}}
\setcounter{footnote}{0}

\newpage


\title{Electroweak Precision Data and Gravitino Dark Matter}

\author{S. Heinemeyer$^1$}
\address{$^1$Instituto de Fisica de Cantabria (CSIC-UC), Santander, Spain}


\pacs{2.0}

\abstract{
Electroweak precision measurements can provide indirect information about
the possible scale of supersymmetry already at the present level of
accuracy.  
We review present day sensitivities of precision data in mSUGRA-type
models with the gravitino as the lightest supersymmetric particle
(LSP). 
The $\chi^2$~fit is based on $\MW$, $\sweff$, $(g-2)_\mu$, $\br(b \to s \ga)$ 
and the lightest MSSM Higgs boson mass, $\Mh$. 
We find indications for relatively light soft supersymmetry-breaking masses,
offering good prospects for the LHC and the ILC, and in some cases
also for the Tevatron. 

}

\maketitle


\section{Introduction}

We have recently analyzed~\cite{ehow3,ehow4} the indications provided
by current experimental data concerning the possible scale of
supersymmetry (SUSY) within the framework of the minimal
supersymmetric extension of the Standard Model (MSSM),
assuming that the soft supersymmetry-breaking scalar masses $m_0$, gaugino
masses $m_{1/2}$ and tri-linear parameters $A_0$ were each constrained to
be universal at the input GUT scale, with the gravitino heavy and the
lightest supersymmetric particle (LSP) being the lightest neutralino
$\neu{1}$ (CMSSM). (For other recent analyses, see \citere{other}.)
However, there are more scenarios for SUSY phenomenology. 
As an example, the gravitino might be the LSP
and constitute the dark matter \cite{ekn} (see also \citere{feng}), a
framework known as the GDM~\cite{GDM}. 

Supersymmetry may provide an important contribution to loop effects that
are rare or forbidden within the Standard Model.
Especially sensitive in this respect are the 
observables $\MW$ and $\sweff$, the loop induced quantities
$(g-2)_\mu$ and $\br(b \to s \ga)$, as well as and the lightest MSSM
Higgs boson mass, $\Mh$, (see \citere{PomssmRep} for a review). 
Another important constraint is provided by 
the cold dark matter (CDM) density $\Omega_{\rm CDM} h^2$ determined
by WMAP and other observations. 
We analyze the precision observables in the context of the GDM,
focusing on parameter combinations that fulfill
$0.094 < \Omega_{\rm CDM} h^2 < 0.129$~\cite{WMAP}.
In order to simplify the analysis in a motivated manner, we
furthermore restrict our attention to scenarios inspired by 
supergravity (mSUGRA), in which the gravitino mass is
constrained to equal $m_0$ at the input GUT scale, and the trilinear
and bilinear soft supersymmetry-breaking parameters are related by
$A_0 = B_0 + m_0$.  In the cases we review here, namely 
$A_0/m_0 = 0, 3/4, 3 - \sqrt{3}, 2$, the regions 
of the $(m_{1/2}, m_0)$ plane allowed by cosmological constraints then take
the form of wedges located at small values of $m_0$~\cite{GDM,eov}.


\section{The \boldmath{$\chi^2$} fit}
\label{sec:chi2fit}

In this Section we review briefly the experimental
data set that has been used for the fits. We focus on parameter points that
yield the correct value of the cold dark matter density, 
$0.094 < \Omega_{\rm CDM} h^2 < 0.129$~\cite{WMAP}, which is, however, not
included in the fit itself. The top quark mass has been fixed to 
$\mt = 172.7 \gev$~\cite{mt1727}, where the experimental uncertainty of 
$\de\mt^{\rm exp} = 2.9 \gev$ has been taken into account in the
parametric uncertainty, see below. 
For the other observables we use the following experimental values
(see \citere{ehow4} and references therein)
\BEA
\MW^{\rm exp} &=& 80.410 \pm 0.032 \gev , \non \\
\sweff^{\rm exp} &=& 0.23153 \pm 0.00016 , \non \\
\amuexp-\amu^{\rm theo,SM} &=& (25.2 \pm 9.2) \times 10^{-10} , \non \\
\br( b \to s \ga ) &=& (3.39^{+ 0.30}_{- 0.27}) \times 10^{-4} .
\EEA
An update to the most recent experimental values would not change our
results in a qualitative manner (see e.g.\ \citere{LCWS05ehow3} for an
analysis of the dependence on $\mt^{\rm exp}$ and $\de\mt^{\rm exp}$). 
For $\Mh$ we use the complete likelihood information available from
LEP~\cite{LEPHiggsSM}. 
Our starting points are the $CL_s(\Mh)$ values provided by the 
final LEP results on the SM Higgs boson search, see Fig.~9 
in~\cite{LEPHiggsSM}, where the $\chi^2$ contribution is obtained by
inversion $CL_s(\Mh)$, see \citere{ehow4} for details. 

Assuming that the five observables listed above are
uncorrelated, a $\chi^2$ fit has been performed with
\BE
\chi^2 \equiv \sum_{n=1}^{4} \KL \frac{R_n^{\rm exp} - R_n^{\rm theo}}
                                 {\si_n} \KR^2 + \chi^2_{\Mh}.
\label{eq:chi2}
\EE
Here $R_n^{\rm exp}$ denotes the experimental central value of the
$n$th observable ($\MW$, $\sweff$, \mbox{$(g-2)_\mu$} and $\br(b \to s \ga)$),
$R_n^{\rm theo}$ is the corresponding GDM prediction and $\si_n$
denotes the combined error (experimental, parametric, intrinsic, 
\citeres{ehow4,PomssmRep}). 
$\chi^2_{\Mh}$ denotes the $\chi^2$ contribution coming from the
lightest MSSM Higgs boson mass as described above.
For details of the theory evaluations see
\citeres{PomssmRep,MWweber,bsg,g-2,feynhiggs} (and references therein).


\section{Results in the GDM}
\label{sec:results}

\begin{figure}[htb!]
\begin{center}
\includegraphics[width=.45\textwidth,height=4.5cm]{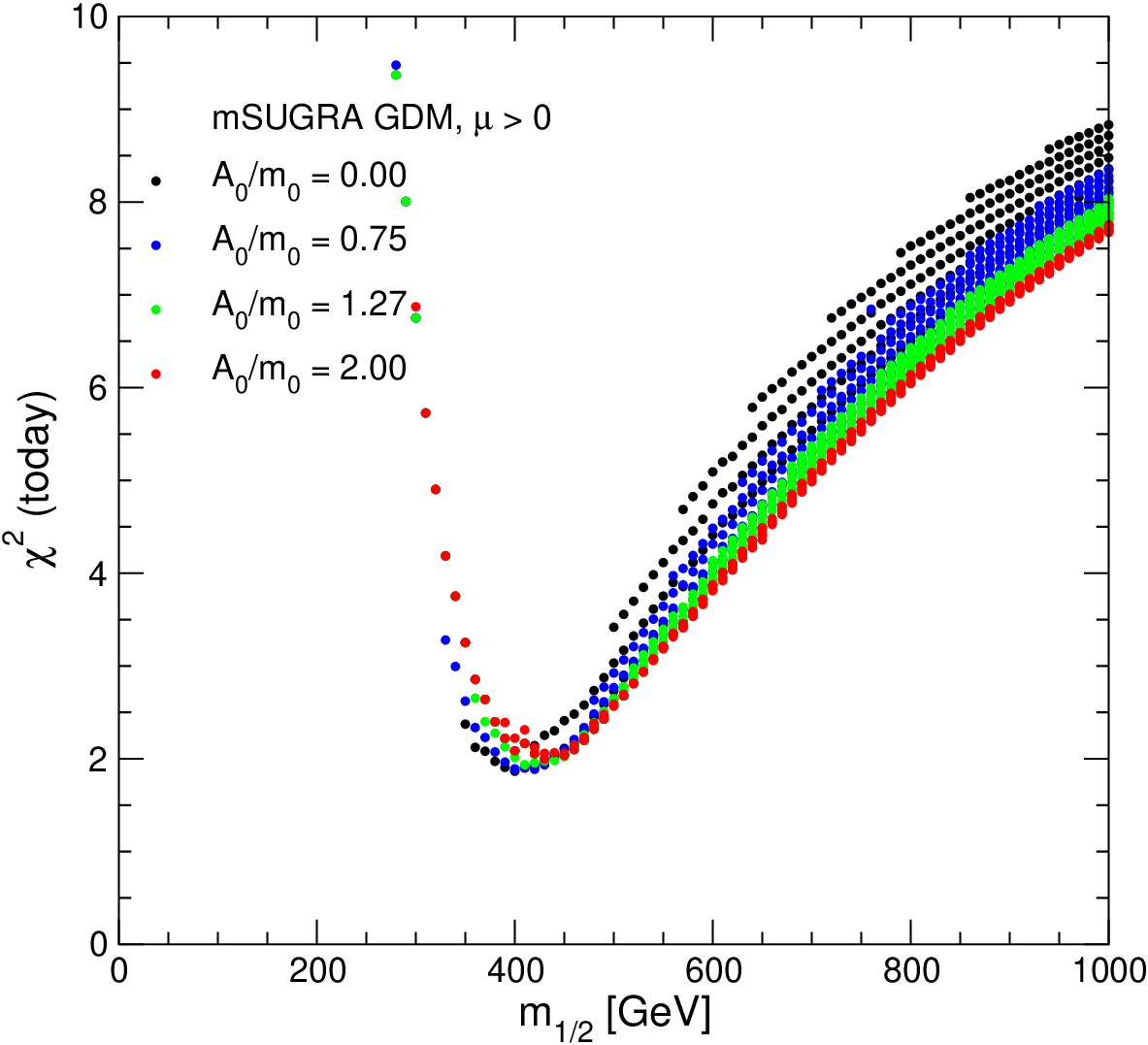}
\includegraphics[width=.45\textwidth,height=4.5cm]{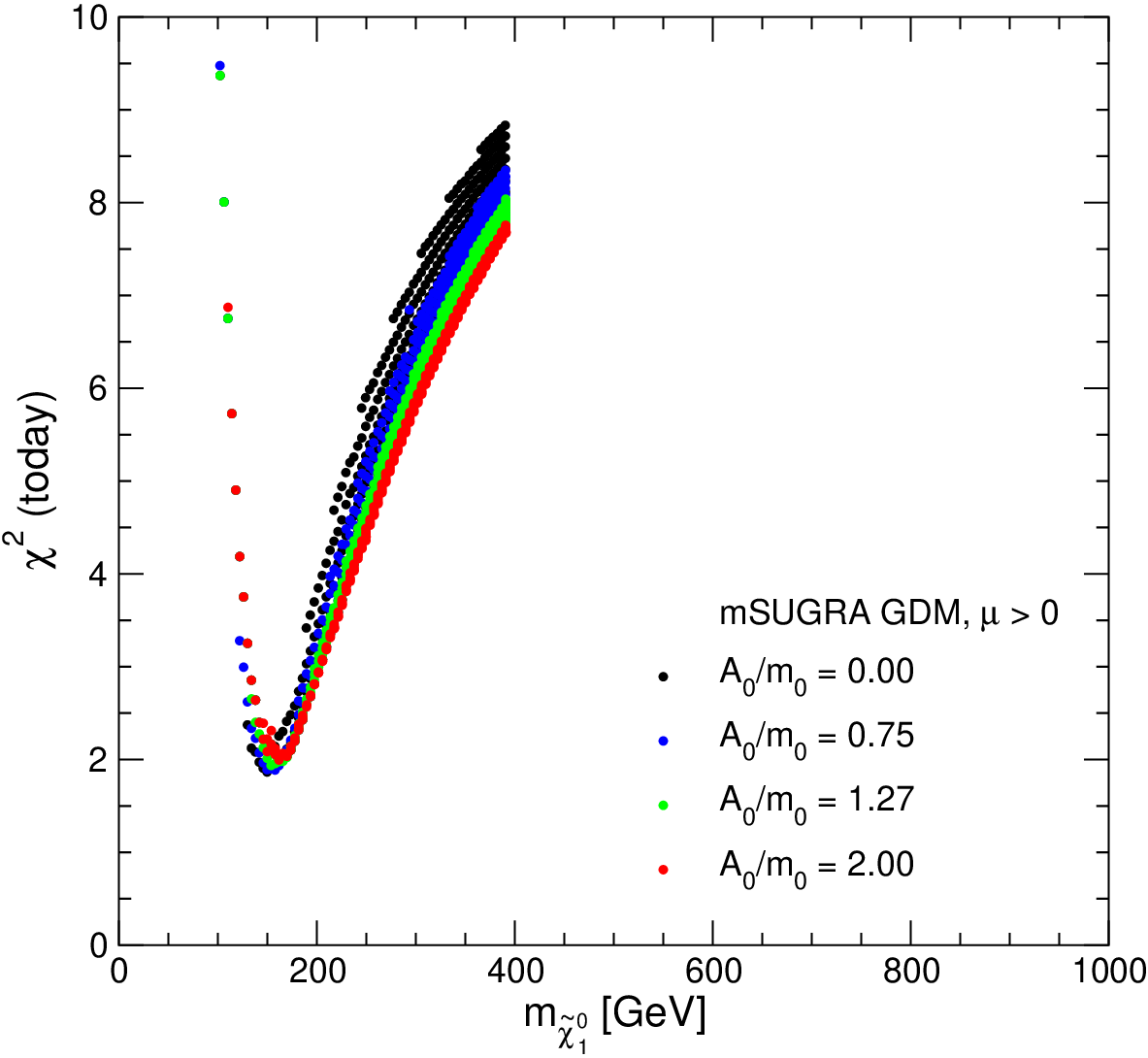}\\[3em]
\includegraphics[width=.45\textwidth,height=4.5cm]{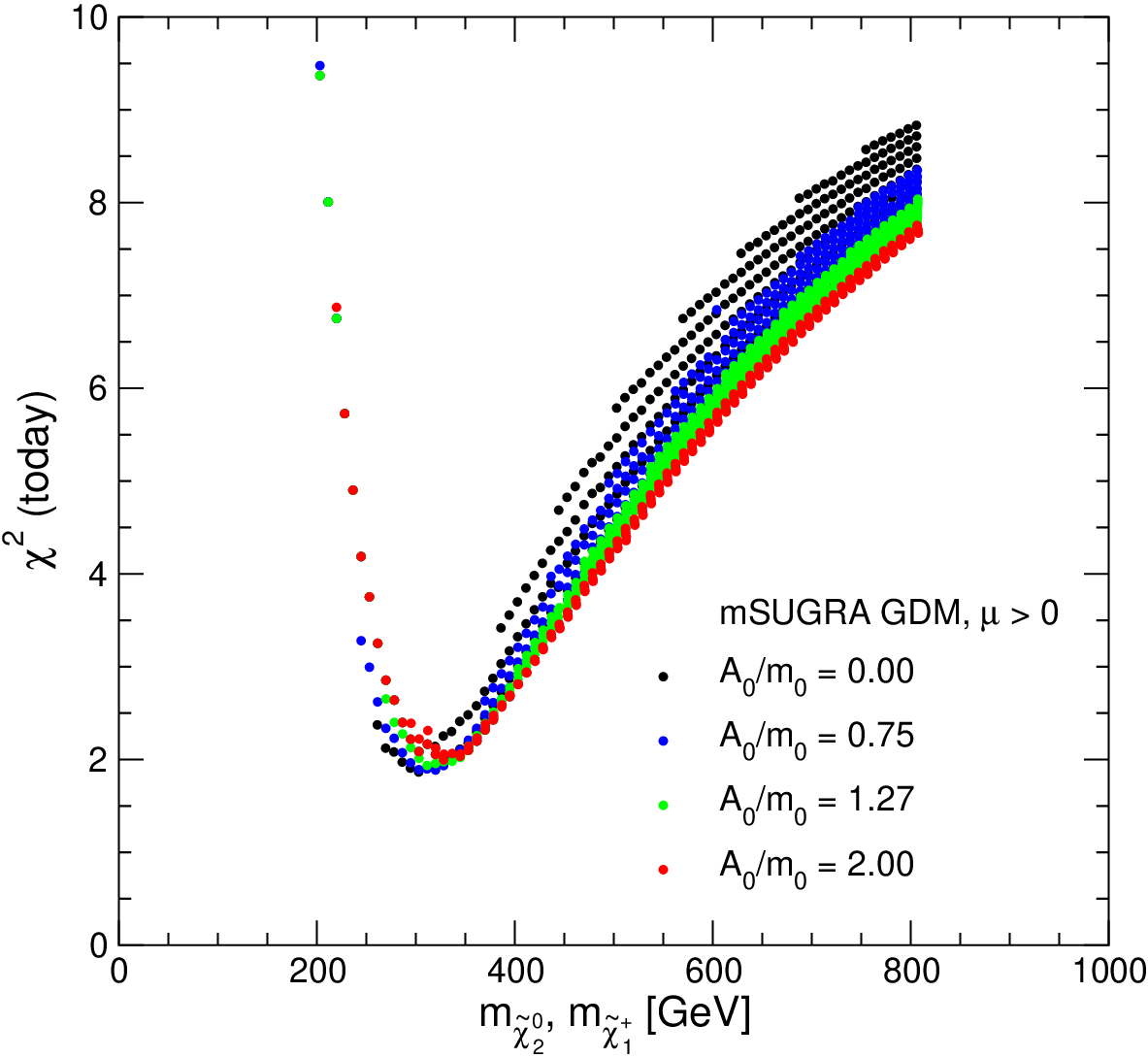}
\includegraphics[width=.45\textwidth,height=4.5cm]{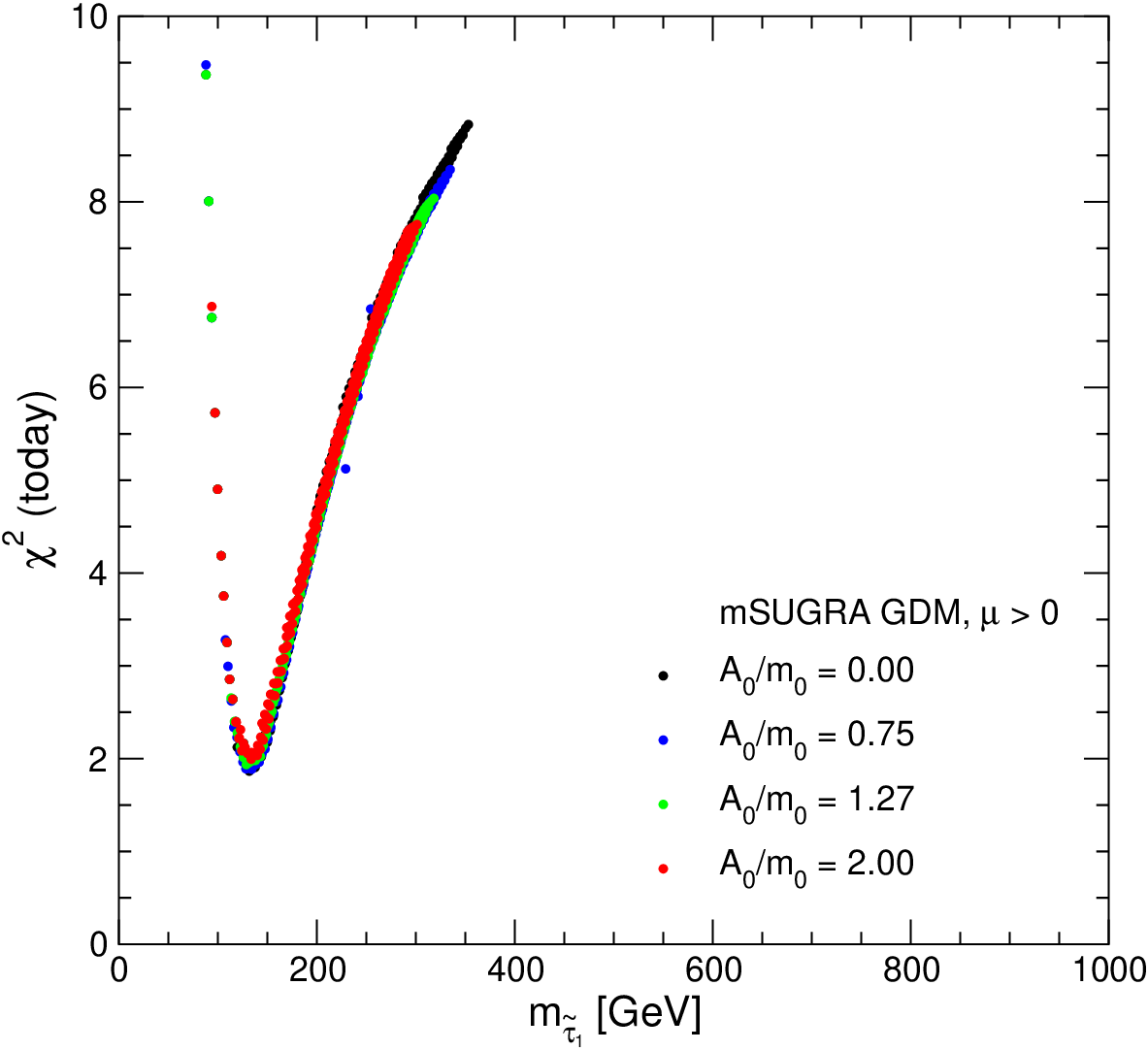}\\[3em]
\includegraphics[width=.45\textwidth,height=4.5cm]{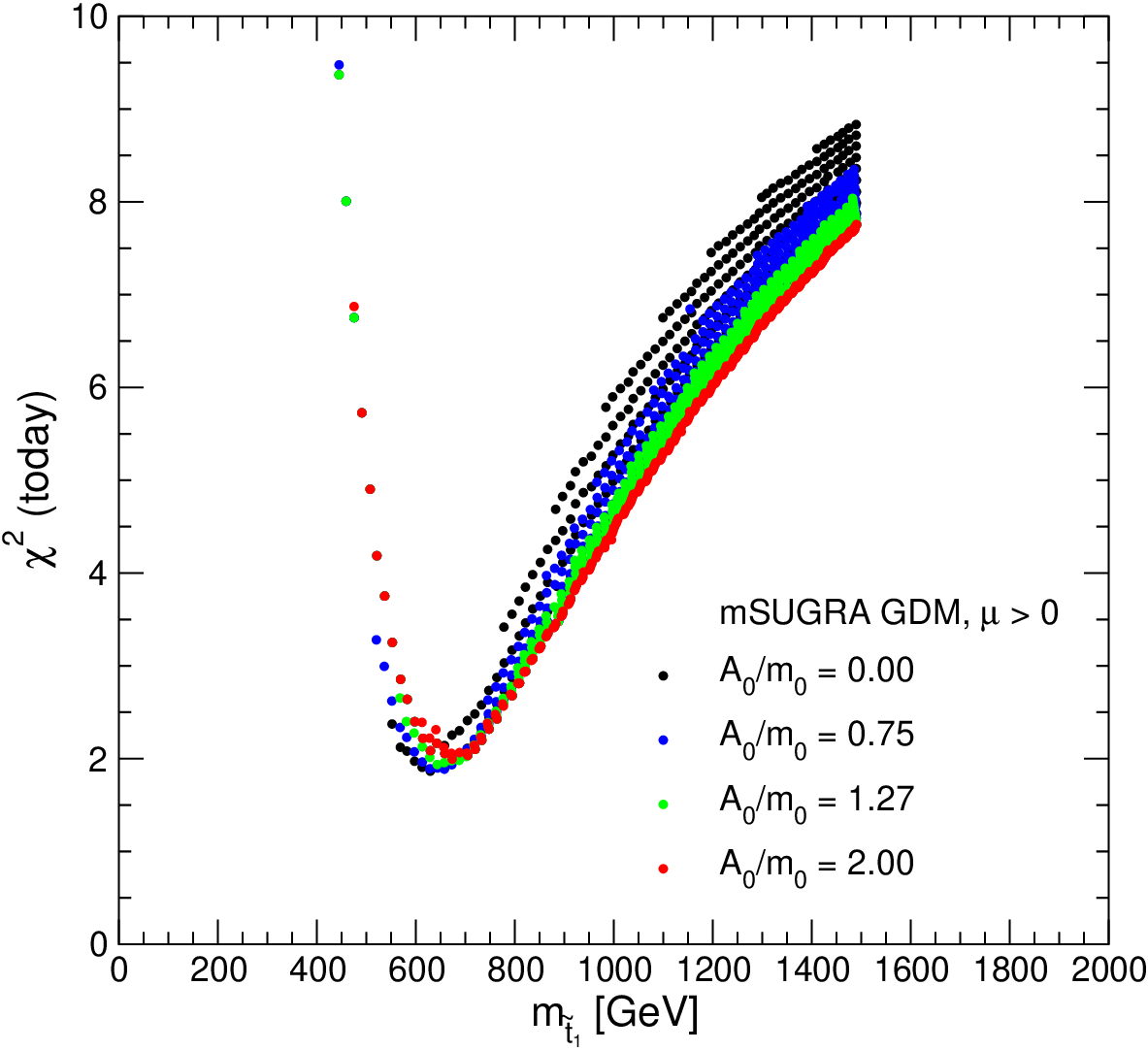}
\includegraphics[width=.45\textwidth,height=4.5cm]{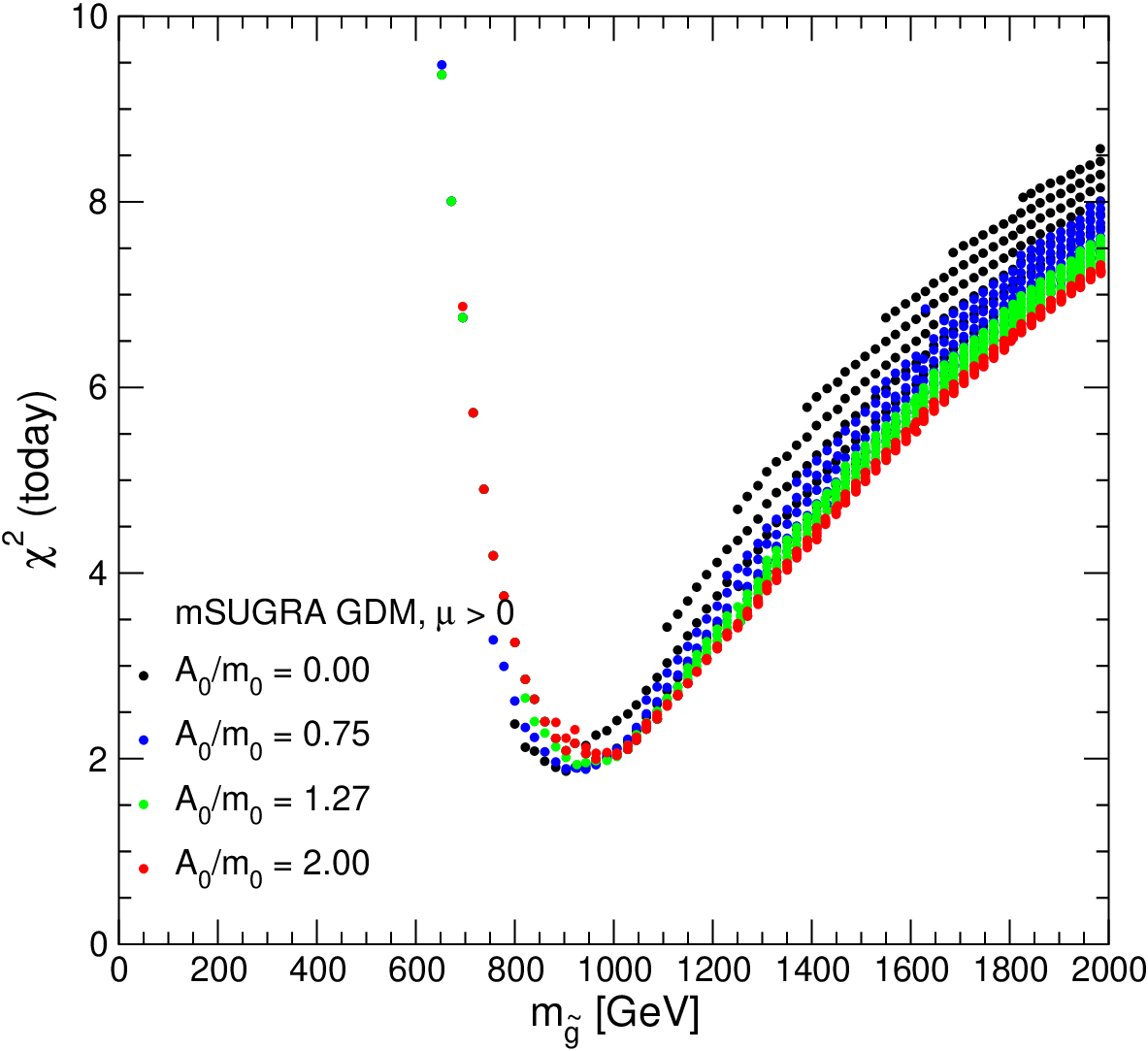}
\caption{
The dependence of the $\chi^2$ function on $m_{1/2}$ for
GDM scenarios with $A_0/m_0 = 0, 0.75, 3 - \sqrt{3}$ and $2$,
scanning the regions where the lighter stau $\Staue$ is the
NLSP, shown as a function of (a) $m_{1/2}$, (b) $\mneu{1}$,
(c) $\mneu{2}$ and $\mcha{1}$, (d) $\mstaue$, (e) 
$\mste$, and (f) $\mgl$.
}
\label{fig:VCMSSMGDM}
\end{center}
\end{figure}

In \reffi{fig:VCMSSMGDM} we show the total $\chi^2$ as a function of
$m_{1/2}$ and various SUSY particle masses. 
The global minimum of $\chi^2$ for
all the GDM models with $A_0/m_0 = 0, 0.75, 3 - \sqrt{3}$ and 2 is
at $m_{1/2} \sim 450 \gev$. However, this minimum is not attained for GDM
models with larger $m_0$, as they do not reach the low-$m_{1/2}$ tip of
the GDM wedge. 
In general, we see in the different panels of \reffi{fig:VCMSSMGDM}
that there might be some hope to observe the lightest $\Stau$ at the
Tevatron, that there are good prospects for observing the $\gl$
and perhaps the $\Stope$ at the LHC, and that the ILC(500) has
good prospects for the $\neu{1}$ and $\Staue$, though these
diminish for larger $m_0$. The ILC(1000) offers much better
chances also for large $m_0$. We recall that, in these GDM scenarios, the
$\Staue$ is the NLSP, and that the $\neu{1}$ is heavier. The
$\Staue$ decays into the gravitino and a $\tau$, and is
metastable with a lifetime that may be measured in hours, days or weeks.

One feature of the class of GDM scenarios discussed here
is that the required value of $\tb$ increases with $m_{1/2}$.  Therefore,
the preference for relatively small $m_{1/2}$ discussed above maps into an
analogous preference for moderate $\tb$, see \citere{ehow4}.
It can be shown that, at the 95 \% confidence level
\begin{equation}
300 \gev \lsim m_{1/2} \lsim 800 \gev, \quad 15 \lsim \tb \lsim 27
\label{GDMlimits}
\end{equation}
in this mSUGRA class of GDM models.\\[-2.5em]




\subsection*{Acknowledgements}

\vspace{-1em}
I thank J.~Ellis, K.A~Olive and G.~Weiglein with whom I derived the
results presented here.\\[-2.5em]


\end{document}